\documentclass[12pt]{iopart}
\usepackage{graphicx}

\begin{document}

\title[The international pulsar timing array project]{The international pulsar timing array project: using pulsars as a gravitational wave detector}

\author{G. Hobbs$^1$, 
A. Archibald$^2$, 
Z. Arzoumanian$^3$, 
D. Backer$^4$, 
M. Bailes$^5$,
N. D. R. Bhat$^5$, 
M. Burgay$^6$, 
S. Burke-Spolaor$^{1,5}$, 
D. Champion$^{1,11}$, 
I. Cognard$^7$, 
W. Coles$^8$,
J. Cordes$^9$,
P. Demorest$^{10}$,  
G. Desvignes$^7$,
R. D. Ferdman$^7$, 
L. Finn$^{24}$,
P. Freire$^{11}$, 
M. Gonzalez$^{12}$, 
J. Hessels$^{13}$, 
A. Hotan$^{14}$,
G. Janssen$^{20}$, 
F. Jenet$^{15}$, 
A. Jessner$^{11}$, 
C. Jordan$^{20}$, 
V. Kaspi$^2$, 
M. Kramer$^{11,20}$, 
V. Kondratiev$^{16}$, 
J. Lazio$^{21}$, 
K. Lazaridis$^{11}$,
K. J. Lee$^{17}$,
Y. Levin$^{18}$, 
A. Lommen$^{19}$,  
D. Lorimer$^{16}$, 
R. Lynch$^{22}$,  
A. Lyne$^{20}$, 
R. Manchester$^1$, 
M. McLaughlin$^{16}$, 
D. Nice$^{23}$, 
S. Oslowski$^{1,5}$, 
M. Pilia$^6$,
A. Possenti$^6$, 
M. Purver$^{20}$, 
S. Ransom$^{10}$, 
J. Reynolds$^{1}$,
S. Sanidas$^{20}$,
J. Sarkissian$^{1}$,
A. Sesana$^{24}$,
R. Shannon$^9$, 
X. Siemens$^{26}$, 
I. Stairs$^{12}$, 
B. Stappers$^{20}$, 
D. Stinebring$^{25}$, 
G. Theureau$^7$, 
R. van Haasteren$^{18}$,
W. van Straten$^5$,
J. P. W. Verbiest$^{16}$, 
D. R. B. Yardley$^{1,27}$,
X. P. You$^{28}$}

\address{$^1$ Australia Telescope National Facility, CSIRO, P.O. Box 76, Epping, NSW 1710 Australia \\
$^2$ Department of Physics, McGill University, Montreal, PQ, H3A 2T8, Canada \\
$^3$ CRESST/USRA, NASA Goddard Space Flight Center, Code 662, Greenbelt, MD 20771, USA\\
$^4$ Astronomy department and radio astronomy laboratory, University of California, Berkeley, CA 94720-3411, USA \\
$^5$ Centre for Astrophysics and Supercomputing, Swinburne University of Technology, P.O. Box 218, Hawthorn VIC 3122, Australia\\
$^6$ Universit\`a di Cagliari, Dipartimento di Fisica, SP Monserrato-Sestu km 0.7, 09042 Monserrato (CA), Italy\\
$^7$ Station de Radioastronomie de Nanay, Observatoire de Paris,
18330 Nan\c cay, France\\
$^8$ Electrical and Computer Engineering, University of California at San Diego, La Jolla, California, USA \\
$^9$ Astronomy Department, Cornell University, Ithaca, NY 14853, USA\\
$^{10}$ National Radio Astronomy Observatory (NRAO), Charlottesville, VA 22903, USA \\
$^{11}$ Max-Planck-Institut f\"{u}r Radioastronomie, Auf Dem H\"{u}gel 69, 53121, Bonn, Germany\\
$^{12}$ Department of Physics and Astronomy, University of British Columbia, 6224 Agricultural Road, Vancouver, BC V6T 1Z1, Canada \\
$^{13}$ Astronomical Institute ÒAnton PannekoekÓ, University of Amsterdam, Kruislaan 403, 1098 SJ Amsterdam, The Netherlands\\
$^{14}$ Department of Imaging and Applied Physics, Curtin University, Bentley, WA, Australia \\
$^{15}$ Center for Gravitational Wave Astronomy, University of Texas at
Brownsville, 80 Fort Brown, Brownsville, TX 78520, USA\\
$^{16}$ Department of Physics, West Virginia University, Morgantown, WV 26506, USA\\
$^{17}$ Department of Astronomy, Peking University, 5 Haidian Lu, Beijing, 100871, China\\
$^{18}$ Leiden University, Leiden Observatory, PO Box 9513, NL-2300 RA Leiden, the Netherlands\\
$^{19}$ Franklin and Marshall College, 415 Harrisburg Pike, Lancaster, PA 17604, USA\\
$^{20}$ Jodrell Bank Centre for Astrophysics, University of Manchester, Manchester, M13 9PL, UK\\
$^{21}$ Naval Research Laboratory, 4555 Overlook Avenue, SW, Washington, DC 20375, USA\\
$^{22}$ Department of Astronomy, University of Virginia, Charlottesville, VA 22904Ð4325, USA\\
$^{23}$ Physics Department, Bryn Mawr College, Bryn Mawr, PA 19010, USA\\
$^{24}$ Center for Gravitational Wave Physics, The Pennsylvania State University, University Park, PA 16802, USA \\
$^{25}$ Physics \& Astronomy Dept., Oberlin College, Oberlin, OH 44074, USA\\
$^{26}$ Center for Gravitation and Cosmology, Department of Physics, University of Wisconsin --- Milwaukee, P.O. Box 413, Wisconsin, 53201, USA\\
$^{27}$ Sydney institute for astronomy, the University of Sydney, NSW, Australia \\
$^{28}$ School of Physical Science and Technology, Southwest University, 2 Tiansheng Road, Chongqing 400715, China \\}
\ead{george.hobbs@csiro.au}
\begin{abstract}
The International Pulsar Timing Array project combines observations of pulsars from both Northern and Southern hemisphere observatories with the main aim of detecting ultra-low frequency ($\sim 10^{-9}-10^{-8}$\,Hz) gravitational waves.  Here we introduce the project, review the methods used to search for gravitational waves emitted from coalescing supermassive binary black-hole 
systems in the centres of merging galaxies and discuss the status of the project.
\end{abstract}

%\maketitle

\section{Introduction}

Since their discovery (Hewish et al. 1968), pulsars have been repeatedly monitored by many large radio telescopes.  The characteristic signature of a radio pulsar is a regular train of pulsed radiation. The intensity, shape and arrival times of the pulses are determined by the physical phenomena causing the emission, the pulsar's magnetosphere, the pulse propagation through the interstellar medium and effects caused by the detection systems at the observatory.  As emphasised throughout this paper, encoded into the pulse arrival times will also be information relating to gravitational-waves (GWs) passing the Earth and/or the pulsar.

Pulsar observations have aided many new discoveries in physics and astronomy.  For instance, the first extra-solar planets were discovered around a pulsar (Wolszczan \& Frail 1992), the first observational evidence for gravitational waves was obtained by studying a pulsar-neutron star binary system (Hulse \& Taylor 1975)\nocite{ht75} and the most stringent tests of general relativity in the strong-field limit come from pulsar observations (Kramer et al. 2006).  These results all relied on a technique known as ``pulsar timing''.  Details of this technique have been described numerous times in the literature (see Lorimer \& Kramer 2005\nocite{lk05} for an overview and Edwards, Hobbs \& Manchester 2006\nocite{ehm06} for full details of the method).  In brief, the observed pulse times-of-arrival (TOAs) are compared with a prediction for the arrival times obtained with a model of the spin, astrometric and orbital parameters of the pulsar and details of the pulse propagation through the interstellar medium.  The deviations between the predicted and the observed TOAs are known as the pulsar `timing residuals'  and indicate unmodelled effects, i.e., $R_i = (\phi_i - N_i)/\nu$ where $\phi_i$ describes the time evolution of the pulse phase based on the model pulse frequency ($\nu$) and its derivatives. $N_i$ is the nearest integer to $\phi_i$. GW signals are not included in a pulsar timing model and, hence, any such waves will induce residuals.  Unfortunately, the expected signal induced by GWs is small, with typical residuals being $<$100\,ns.

The TOA precision achievable for the majority of pulsars is $\sim 1$\,ms and most pulsars show long-term timing irregularities that would make the detection of the expected GW signal difficult or impossible (e.g. Hobbs, Lyne \& Kramer 2006\nocite{hlk06}).  However,  a sub-set of the pulsar population, the millisecond pulsars, have very high spin rates, much smaller timing irregularities and can be observed with much greater TOA precision. Recent observations of PSR~J0437$-$4715 have shown that TOA precisions of $\sim 30$\,ns can be achieved (see \S \ref{sec:status}) and over 10\,yr the root-mean-square (rms) timing residuals are 200\,ns (Verbiest et al. 2008)\nocite{vbv+08}. 

In \S2 of this paper we describe the induced timing residuals caused by GWs. The expected sources of detectable GW signals are given in \S3.  We summarise the International Pulsar Timing Array project in \S4 and highlight future telescopes and timing array projects in \S5.

% Discuss pulsars - types of pulsars - why are msps so good}
% Basic explanation of pulsar timing

% plot ppdot diagram indicating IPTA pulsars

\section{Induced timing residuals caused by gravitational waves}

Sazhin (1978) and Detweiler (1979) first showed that a GW signal causes a fluctuation in the observed pulse frequency $\delta \nu/\nu$ which affects the pulsar timing residuals at time $t$ from the initial observation as
\begin{equation}
R(t) = -\int_0^t \frac{\delta \nu(t)}{\nu} dt.
\end{equation}
The Doppler shift can be shown to have the form
\begin{equation}
\frac{\delta \nu}{\nu} = H^{ij}(h_{ij}^e - h_{ij}^p)
\end{equation}
where $h_{ij}^e$ is the GW strain at the Earth at the time of observation, $h_{ij}^p$ the strain at the pulsar when the electromagnetic pulse was emitted (typically $\sim 1000$\,yr ago) and $H^{ij}$ is a geometrical term that depends upon the angle between the Earth, pulsar and GW source.  This equation was derived assuming a plane gravitational wave and is accurate to first order in $h_{ij}$ for all GW wavelengths. Note, this expression holds even if the wave is not sinusoidal.  Full details of the exact form of the induced residuals are given by Hobbs et al. (2009a). Standard pulsar timing techniques absorb any low-frequency GWs by fitting for the pulsar's spin-down and so the time span of the data provides a lower bound on the GW frequencies that are detectable (currently $> 10^{-9}$\,Hz).  Pulsars are typically only observed once every few weeks and so the data sampling limits the maximum detectable GW frequency to $\sim 10^{-7}$\,Hz.

\begin{figure}
\begin{center}
\includegraphics[width=6cm,angle=-90]{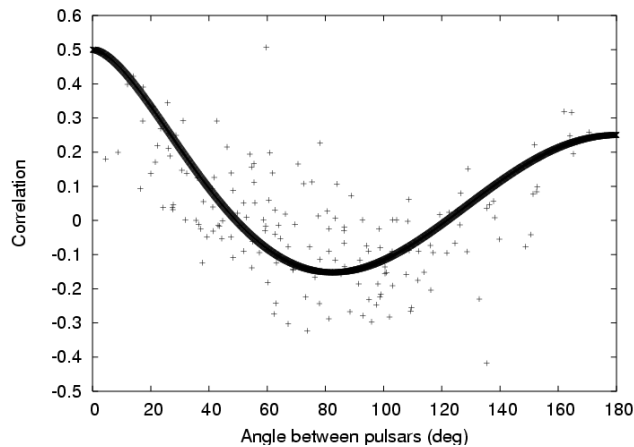}
\caption{The expected correlation in the timing residuals of pairs of pulsars as a function of angular separation for an isotropic GW background.}\label{fg:hdCurve}
\end{center}
\end{figure}

It is not possible to determine the exact origin of the timing residuals with a single pulsar data set.  For instance, residuals may be caused by irregularities in terrestrial time standards, errors in the planetary ephemeris, irregular spin-down of the pulsar, calibration effects or GWs.  These different effects may be distinguished by searching for correlations in the residuals of many pulsars.  Residuals caused by the irregular spin-down of one pulsar will be uncorrelated with the residuals observed for a different pulsar. If there are irregularities in terrestrial time standards then the residuals for all pulsars will be correlated.  For an isotropic, stochastic GW background, the GW strain at each pulsar will be uncorrelated, but the GW strain at the Earth provides a common signal.  The expected correlation as a function of angular separation between two pulsars was determined by Hellings \& Downs (1983)\nocite{hd83} and is shown in Figure~\ref{fg:hdCurve}. Note, this figure shows the expected function for GWs described by general relativity.  Lee, Jenet \& Price (2008)\nocite{ljp08} produced similar angular correlation functions for more general theories of gravity.

% Summary of HD-curve etc.

\section{Expected sources of gravitational waves}

Expected sources of GWs in the pulsar frequency band can be divided into 1) single persistent sources, 2) individual burst sources and 3) stochastic GW backgrounds.   Here we will only consider the GW emission from coalescing supermassive binary black-hole systems in the centres of merging galaxies.  Details on the GW emission from other sources such as cosmic strings or from the inflationary era can be found in Maggiore (2000)\nocite{mag00}.

\begin{figure}
\begin{center}
\includegraphics[width=6cm,angle=-90]{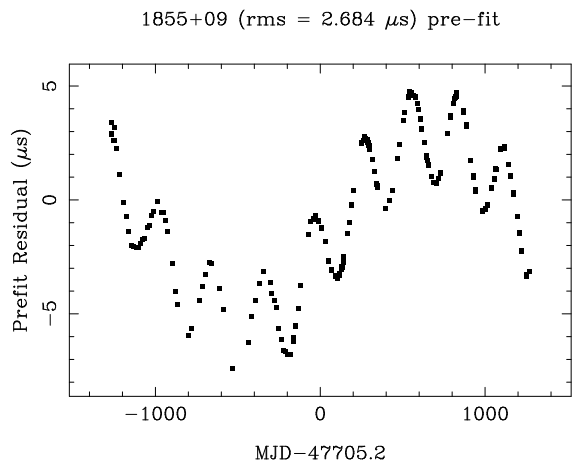}
\caption{Simulation of the induced timing residuals for PSR B1855$+$09 caused by a postulated supermassive binary black hole system in the radio galaxy 3C66B.}\label{fg:3c66b}
\end{center}
\end{figure}

Sudou et al. (2003)\nocite{simt03} obtained observational evidence for a coalescing supermassive black-hole binary system at the centre of the radio galaxy 3C66B.  Jenet et al. (2004)\nocite{jllw04} calculated the expected induced timing residuals from such a system (see Figure~\ref{fg:3c66b}) and, by comparison with actual pulsar data (Kaspi et al. 1994), were able to rule out the existence of the postulated system with 95\% confidence.  Similarly, the blazar OJ287 has an observed periodicity of $\sim 12$\,yr which can be modelled as a binary black-hole system (e.g. Valtonen et al. 2009)\nocite{vnv+09}. The most recent model parameters for this system suggest that the induced timing residuals will lie around $\sim 1-10$\,ns and would therefore be undetectable in existing data sets.

Using models of black-hole masses and merging rates, Sesana, Vecchio \& Colacino (2008)\nocite{svc08} and Sesana, Vecchio \& Volonteri (2009)\nocite{svv09} studied theoretically the detectability of GW signals from coalescing black-hole systems.  They showed that it is unlikely that an individual source of GWs would be detectable at current PTA sensitivities (rms timing residuals $\sim 10$\,ns are required to detect such individual sources), but that the amplitude of a stochastic background created by massive (10$^9$\,M$_\odot$) black-hole binary systems in distant ($z \sim 2$) galaxies could be just below current observational limits.

The current best limit (Jenet et al. 2006) used observations from the Parkes and Arecibo observatories. The characteristic strain spectrum of an isotropic, stochastic background as\footnote{see Sesana et al. (2008) for a more accurate model} can be written as $h_c(f) = A(f/f_{\rm 1yr})^{-2/3}$. The most recent limit gave $A < 10^{-14}$ (where $f_{\rm 1yr} = 1/[1{\rm yr}]$).  New techniques to obtain limits (e.g. van Haasteren et al. 2009, Anholm et al. 2008)\nocite{an08}\nocite{vlml09} and new data sets (see below) should significantly improve on this limit in the near future.

\section{Status of the IPTA project}\label{sec:status}

\begin{figure}
\begin{center}
\includegraphics[width=6cm,angle=-90]{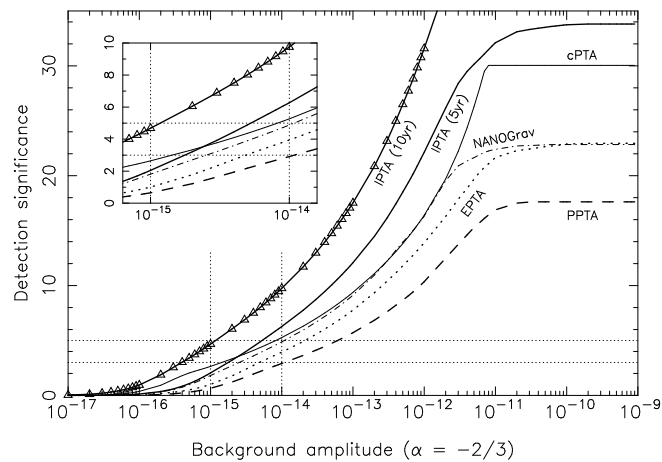}
\caption{Detection significance of a GW background with given amplitude for data spanning five years for different pulsar timing arrays.  Also shown is the expected detection significance for the IPTA project after 10 years.  The detection significance is defined in the appendix of Verbiest et al. (2009).}\label{fg:ipta_sens}
\end{center}
\end{figure}

The International Pulsar Timing Array (IPTA) project is a collaboration between three individual projects. The European project (EPTA; Janssen et al. 2008\nocite{jsk+08b}) currently combines data from four European telescopes (Effelsberg, Jodrell Bank, Nan\c cay and Westerbork -- a new telescope in Sardinia is currently being commissoned). The North American project (NANOGrav; Jenet et al. 2009\nocite{jen09}) observes with the Arecibo and Green Bank telescopes. The Parkes project (PPTA; e.g. Hobbs et al. 2009b\nocite{hbb+09}) uses the Parkes radio telescope. Basic parameters of the pulsars observed are listed in Table~\ref{tb:psrs}.  

Some of our most precise timing residuals have been obtained for PSR~J0437$-$4715. The rms timing residuals over $\sim$1\,yr obtained with the Parkes radio telescope at an observing frequency of $\sim 3$\,GHz is $\sim 60$\,ns and some individual TOA uncertainties are $\sim 30$\,ns.  However, it is not yet clear whether the low rms of the residuals will continue with longer data spans.  Verbiest et al. (2008) presented observations of this pulsar over the previous 10\,yr and detected the presence of low-frequency timing irregularities.  These irregularities may limit the rms achievable over many years.  More recently Verbiest et al. (2009) studied the timing stability for a sample of 20 millisecond pulsars over time scales of up to $10$\,yr.  Even though  PSR~J1939$+$2134 showed significant unmodelled residuals that may make this pulsar unusable for long-term GW detection experiments, the timing of most of the other pulsars was shown to be stable enough for GW detection over decadal time scales.

For the pulsars in Table~\ref{tb:psrs} we have used typical, current signal-to-noise (S/N) measurements of pulse profiles obtained at each of the IPTA observatories to calculate the best possible timing precision given the period of the pulsar and its pulse shape.  We emphasise that the actual rms timing residuals currently being obtained are usually significantly worse. Out of 37 pulsars, thirteen have expected TOA uncertainties $< 250$\,ns, seven between $250$ and $500$\,ns, a further nine between $500$\,ns and $1 \mu$s, five between $1$ and $2 \mu$s and the remaining three $> 2 \mu$s.  For the case of a stochastic GW background created by coalescing supermassive black-hole systems, we have calculated the signal-to-noise ratio of a measurement of the GWB amplitude for the various PTAs. Our analysis is based on that presented by Verbiest et al. (2009) who use the term "detection significance" for this signal-to-noise ratio.  The results are plotted in Figure 3. Here we have assumed that 1) each project obtains approximately one observation every three weeks over five years and 2) the timing residuals (with no simulated GWs signals present) are ÔwhiteÕ (i.e., they have a flat spectrum)  with an rms given by the expected TOA uncertainties (i.e., all timing noise, ISM effects and calibration errors are assumed to be negligible). We know that some of the pulsars in Table 1 exhibit non-white timing noise which would degrade the sensitivity on a 5\,yr scale significantly and would be even more detrimental on a 10\,yr scale. However, we do not have this information for all of the pulsars in Table 1, so we cannot uniformly improve the analysis. Our result represents ``the best we could do with current systems''. There is some reason to hope that timing noise can be modeled and corrected (Lyne et al. in preparation). If the background amplitude is just below current upper bounds (Jenet et al. 2006)\nocite{jhv+06} then such a five-year data set would produce a signal-to-noise ratio of $\sim 6$.  However, if the amplitude is closer to $A = 10^{-15}$ then the background will be barely detectable.  For similar data sets, but with data spanning 10\,yr, a background with $A = 10^{-15}$ will still only have a signal-to-noise ratio of $\sim 5$ (triangle symbols).  Also shown on this plot are the detection significance obtained using the PPTA, NANOGrav and EPTA data individually.  Clearly, a detection of the GW background will require the combination of these data sets.  Finally, the thin solid line (marked `cPTA') indicates a canonical timing array where 20 pulsars are each timed for 100\,ns over 5\,yr with weekly sampling.

\begin{figure*}
\begin{center}
\includegraphics[width=6cm,angle=-90]{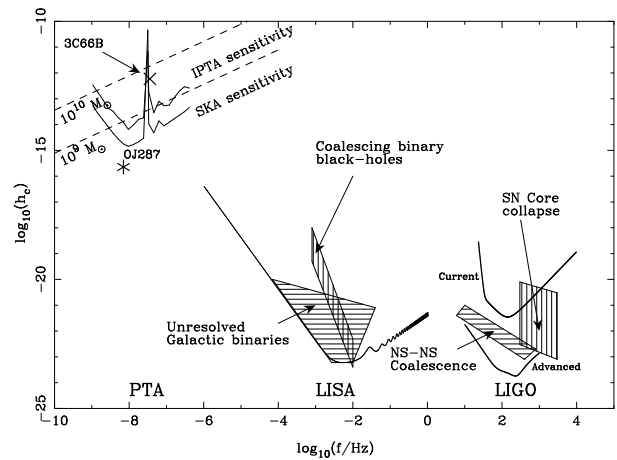}
\caption{The sensitivity to individual sources of GWs is shown for the IPTA and a possible future experiment with the SKA. The expected signals from coalescing black holes at the cores of 3C66B and OJ287 are shown.  The dashed lines indicate the expected signals from black hole binary systems with chirp masses of $10^9$ and $10^{10}$\,M$_\odot$ respectively, situated in the Virgo cluster. For comparison, sensitivity curves of LIGO and LISA and predicted signal levels are also shown.}\label{fg:sens}
\end{center}
\end{figure*}

In Figure~\ref{fg:sens}, we show the sensitivity of the IPTA project to single, persistent sources of GWs (Yardley et al., in preparation). This figure indicates that sources similar to the postulated binary system in the radio galaxy 3C66B would be clearly detectable with any polarisation and at any position in the sky. 

 A significant reduction in the required time to detect GWs would be made if the rms timing residuals could be further reduced.  Techniques that may achieve such a reduction include 1) observing at higher frequencies\footnote{Note that for some pulsars improved TOA uncertainties can be obtained from low frequency observations. However, compared with higher frequency data such observations are more affected by interstellar medium propagation delays which add uncorrelated timing noise.}, 2) discovering new millisecond pulsars, 3) obtaining pulse TOAs through use of polarisation information (van Straten 2006)\nocite{van02}, 4) increasing bandwidth and/or observing time and 5) observing only during periods when the pulsar is bright because of scintillation. Also note that the simulations described above assume no previous data on each pulsar.  Most of these pulsars have been observed for years to decades with various observing systems.  The use of these earlier data sets will improve our sensitivity to GW signals.

% Figure plotting the sky position of all the pulsars
% Give current limits and future possibilities.

%Include sentence or two about outreach activities
%In order to describe our project to the general community and to begin to train the next generation of GW astronomers, some of the observations at Parkes are carried out by high-school students as part of the PULSE@Parkes project (Hobbs et al., in press).

%NB 0711 gves 0.4 with Parkes with best profile, but not with typical profile
\begin{table*}
\caption{Pulsars observed as part of the International Pulsar Timing Array project. The best possible rms values for observations (with current observing systems) from the PPTA, EPTA and NANOGrav projects are estimated in the final three columns.}\label{tb:psrs}
\begin{tiny}\begin{center}
\begin{tabular}{llllrllll}\hline
PSR B & PSR J & P & $P_b$ & S1400 & Array & PPTA & EPTA & NANOGrav \\
           & & (ms) & (d) & (mJy) & & ($\mu$s)& ($\mu$s)& ($\mu$s)  \\ \hline
- & J0030$+$0451 & 4.87 & - & 0.6 & EPTA, NANOGrav & - & 0.54 & 0.31 \\
- & J0218$+$4232 & 2.32 & 2.03 & 0.9 & NANOGrav & - & - & 4.81\\
- & J0437$-$4715 & 5.76 & 5.74 & 142.0 & PPTA &  0.03 & - & - \\
- & J0613$-$0200 & 3.06 & 1.20 & 1.4 & PPTA, EPTA, NANOGrav &  0.71 &  0.45 & 0.50\\
 & J0621$+$1002 & 28.85 & 8.32& 1.9& EPTA & - & 9.58 & - \\ \\
 
- & J0711$-$6830 &  5.49 & -   &   1.6     & PPTA &  1.32 & - & - \\ 
 - & J0751$+$1807 & 3.48 & 0.3 & 3.2 & EPTA & - & 0.78 & - \\
-  & J0900$-$3144 & 11.1 & 18.7 & 3.8 & EPTA & - & 1.55 & - \\
  - & J1012$+$5307 &  5.26 & 0.60  & 3.0 & EPTA, NANOGrav & - & 0.32 & 0.61\\
- & J1022$+$1001 &16.45 & 7.81 & 3.0 & PPTA, EPTA & 0.37 & 0.48 & - \\ \\

- & J1024$-$0719 & 5.16 & -      & 0.7    & PPTA, EPTA  & 0.43 & 0.25 & -\\ 
- & J1045$-$4509 & 7.47 & 4.08  & 3.0        & PPTA & 2.68 & - & -  \\
- & J1455$-$3330 & 7.99 & 76.17 & 1.2 & EPTA, NANOGrav & - & 3.83 & 1.60\\ 
- & J1600$-$3053 & 3.60 & 14.35   & 3.2       & EPTA, PPTA & 0.32 &  0.23 & - \\
- & J1603$-$7202 & 14.84  & 6.31  & 3.0        & PPTA & 0.70 & - & -  \\ \\

- & J1640$+$2224 & 3.16 & 175.46 & 2.0 & EPTA, NANOGrav & - & 0.45 & 0.19\\ 
- & J1643$-$1224 &  4.62       &  147.02   & 4.8      & PPTA, EPTA, NANOGrav & 0.57 & 0.56 & 0.53\\
- & J1713$+$0747 & 4.57       &  67.83   & 8.0      & PPTA, EPTA, NANOGrav & 0.15 & 0.07 & 0.04 \\ 
- & J1730$-$2304 & 8.12        &  -   & 4.0      & PPTA, EPTA & 0.83 & 1.01 & - \\
- & J1732$-$5049 & 5.31        &  5.26   & -      & PPTA &  1.74 & - & - \\ \\

- & J1738$+$0333 & 5.85 & 0.35 & - & NANOGrav & - & - & 0.24  \\
- & J1741$+$1351 & 3.75 & 16.34 & - & NANOGrav & - & - & 0.19 \\
- & J1744$-$1134 & 4.08        &   -      & 3.0 & PPTA, EPTA, NANOGrav & 0.21 & 0.14 & 0.14\\
- & J1751$-$2857 & 3.91 & 110.7 & 0.06 & EPTA & - & 0.90 & - \\  
B1821$-$24 & J1824$-$2452 & 3.05        &   -       & 0.2 & PPTA, EPTA & 0.39 & 0.24 & -\\ \\

- & J1853$+$1303 & 4.09 & 115.65 & 0.4 & NANOGrav &-  & -& 0.17 \\ 
B1855$+$09 & J1857$+$0943 & 5.37       &   12.33     &   5.0 & PPTA, EPTA, NANOGrav & 0.82 & 0.44 & 0.25 \\
- & J1909$-$3744 &  2.95       &   1.53 & 3.0       & PPTA, EPTA, NANOGrav & 0.19 & 0.04 & 0.15\\ 
- & J1910$+$1256 & 4.98 & 58.47 & 0.5&  EPTA, NANOGrav & - & 0.99 & 0.17 \\ 
- & J1918$-$0642 &  7.65 & 10.91   & -        & EPTA, NANOGrav & - & 0.87 & 1.08\\ \\

B1937$+$21 & J1939$+$2134 & 1.56       &    -    &10.0    & PPTA, EPTA, NANOGrav & 0.11 & 0.02 & 0.03 \\ 
B1953$+$29 & J1955$+$2908 & 6.13 & 117.35 & 1.1 & NANOGrav & - & - & 0.18 \\
- & J2019$+$2425 &  3.94 & 76.51     & -       & NANOGrav & - & - & 0.66 \\  
- & J2124$-$3358 & 4.93        &   -   & 1.6     & PPTA & 1.52 & - & -\\ 
- & J2129$-$5721 & 3.73         &  6.63  & 1.4      & PPTA &  0.87 & -  & -\\ \\

- & J2145$-$0750 & 16.05        &  6.84   & 8.0      & PPTA, EPTA, NANOGrav & 0.86 & 0.40 & 1.37\\ 
 - & J2317$+$1439 & 3.44        & 2.46    & 4.0        & NANOGrav &- & 0.81 & 0.25\\

\hline
\end{tabular}
\end{center}\end{tiny}
\end{table*}
% Perhaps receiver noise in one hour observation

%\begin{figure*}
%\begin{center}
%\includegraphics[width=7cm,angle=-90]{0437_res.ps}
%\caption{Timing residuals for PSR~J0437$-$4715 obtained using a digital filterbank system at the Parkes radio telescope.}\label{fg:0437_res}
%\end{center}
%\end{figure*}

% Discovery of new pulsars - mention ARCC and other projects

\section{Future data sets}

Current pulsar timing array projects have the potential to make a detection of a GW background or a single GW source.  Unfortunately, the predicted S/N of any such detection is too low for detailed studies of the GW properties.  A high S/N detection of a background would provide a test of general relativity (Lee et al. 2009) and allow a detailed understanding of the properties of the sources that form the GW background.

Within $\sim$10\,yr  it is expected that various new telescopes will be able to improve the existing timing projects. For instance, the Australian Square Kilometre Array Pathfinder (ASKAP) and the South African Karoo Array Telescope (MeerKAT) will increase the number of known pulsars and provide a larger number of telescopes in the Southern Hemisphere for high-precision pulsar timing (Johnston et al. 2007\nocite{jtb+07}, http://www.ska.ac.za/).  It is also likely that a new 500\,m diameter telescope (FAST) will be built in China by 2014 (Nan et al. 2006)\nocite{nwz+06}.  This telescope will be an ideal instrument for pulsar searching and timing and should significantly improve our timing precision for a large number of pulsars.  The EPTA is currently undertaking a plan to create the Large European Array for Pulsars (LEAP).  The LEAP project will coherently combine the signals from the five major European observatories to create a telescope that will have a sensitivity similar to that of the illuminated Arecibo telescope.

On a slightly longer timescale, the Square Kilometre Array (SKA) telescope is planned (current plans are for the full SKA to be completed in 2022, with initial observations with a 10\% SKA to begin in 2018). The huge collecting area of this telescope implies that it will be able to time many hundreds of pulsars at the level currently achieved for a few pulsars (Kramer et al. 2004).  Such timing data sets will revolutionise GW detection and analysis projects by enabling high S/N detections of GW sources.

\section{Outreach}

Many of the IPTA projects have components that are being used to train the first generation of GW astronomers and to discover new pulsars useful for timing array projects.   Arecibo pulsar survey data are currently being obtained and analysed both by high school students in the USA (through the Arecibo Remote Command Center) and the public Worldwide as part of the Einstein@Home project\footnote{http://arcc.phys.utb.edu/ARCC, http://einstein.phys.uwm.edu/}. Students in West Virginia, U.S.A., are able to search for pulsars using observations taken with the Green Bank telescope\footnote{http://www.pulsarsearchcollaboratory.com}.  The PULSE@Parkes project (Hollow et al. 2008) allows students in Australia and overseas to carry out timing observations for the PPTA project (Hobbs et al. 2009c). The Mid-Atlantic Relativistic Initiative in Education (MARIE) is creating tools for introducing gravitational waves in high school and college courses\footnote{ http://www.fandm.edu/marie}.

\section{Conclusion}

Given current theoretical models, it is likely that ultra-low frequency GWs will be detected by pulsar timing experiments within 5-10\,yr.  The first detections are expected to be of an isotropic, stochastic GW background created by coalescing supermassive binary black-hole systems.    It is expected that GW astronomy using pulsars will become commonplace in the SKA era during which the background will be analysed in detail and individual binary black-hole systems and burst GW sources will be detectable.

\subsection{Acknowledgements}

We acknowledge the dedication and skills of the engineers and support staff at the various observatories without whom the IPTA project could not exist.

%\bibliographystyle{aipproc} 
%\bibliography{journals,modrefs,psrrefs,crossrefs}

\end{document}